\documentclass[aps,prx,twocolumn,superscriptaddress,secnum]{revtex4-2}

\usepackage{graphicx}
\usepackage[version=3]{mhchem} 
\usepackage[T1]{fontenc}       
\usepackage{verbatim}
\usepackage[utf8]{inputenc}
\usepackage{siunitx}
\usepackage{float}
\usepackage{comment}
\usepackage{textgreek}
\usepackage{sidecap} 

\usepackage{xcolor}

\begin{document}



\title{Scattering makes a difference in circular dichroic angle-resolved photoemission}

\author{Honey Boban}
\affiliation{Peter Gr\"unberg Institut (PGI-6), Forschungszentrum J{\"u}lich GmbH,
52428 J{\"u}lich, Germany}
\affiliation{JARA-FIT (Fundamentals of Future Information Technology), J\"ulich-Aachen Research Alliance, Forschungszentrum J\"ulich and RWTH Aachen University, 52425 Jülich, Germany}

\author{Mohammed Qahosh}
\affiliation{Peter Gr\"unberg Institut (PGI-6), Forschungszentrum J{\"u}lich GmbH,
52428 J{\"u}lich, Germany}

\author{Xiao Hou}
\affiliation{Peter Gr\"unberg Institut (PGI-6), Forschungszentrum J{\"u}lich GmbH,
52428 J{\"u}lich, Germany}
\affiliation{JARA-FIT (Fundamentals of Future Information Technology), J\"ulich-Aachen Research Alliance, Forschungszentrum J\"ulich and RWTH Aachen University, 52425 Jülich, Germany}


\author{Tomasz Sobol}
\affiliation{National Synchrotron Radiation Centre SOLARIS, Jagiellonian University, Krak\'ow, Poland}

\author{Edyta Beyer}
\affiliation{National Synchrotron Radiation Centre SOLARIS, Jagiellonian University, Krak\'ow, Poland}

\author{Magdalena Szczepanik}
\affiliation{National Synchrotron Radiation Centre SOLARIS, Jagiellonian University, Krak\'ow, Poland}

\author{Daniel Baranowski}
\affiliation{Peter Gr\"unberg Institut (PGI-6), Forschungszentrum J{\"u}lich GmbH,
52428 J{\"u}lich, Germany}

\author{Simone Mearini}
\affiliation{Peter Gr\"unberg Institut (PGI-6), Forschungszentrum J{\"u}lich GmbH,
52428 J{\"u}lich, Germany}

\author{Vitaliy Feyer}
\affiliation{Peter Gr\"unberg Institut (PGI-6), Forschungszentrum J{\"u}lich GmbH,
52428 J{\"u}lich, Germany}

\author{Yuriy Mokrousov}
\affiliation{Peter Gr\"unberg Institut (PGI-1), Forschungszentrum J{\"u}lich GmbH,
52428 J{\"u}lich, Germany}
\affiliation{Institute of Physics, Johannes Gutenberg University Mainz, 55099 Mainz, Germany}

\author{Keda Jin}
\affiliation{Peter Gr\"unberg Institut (PGI-3), Forschungszentrum J{\"u}lich GmbH,
52428 J{\"u}lich, Germany}
\affiliation{JARA-FIT (Fundamentals of Future Information Technology), J\"ulich-Aachen Research Alliance, Forschungszentrum J\"ulich and RWTH Aachen University, 52425 Jülich, Germany}
\affiliation{Institute for Experimental Physics II B, RWTH Aachen University, 52074 Aachen, Germany}

\author{Tobias Wichmann}
\affiliation{Peter Gr\"unberg Institut (PGI-3), Forschungszentrum J{\"u}lich GmbH,
52428 J{\"u}lich, Germany}
\affiliation{JARA-FIT (Fundamentals of Future Information Technology), J\"ulich-Aachen Research Alliance, Forschungszentrum J\"ulich and RWTH Aachen University, 52425 Jülich, Germany}
\affiliation{Institute for Experimental Physics IV A, RWTH Aachen University, 52074 Aachen, Germany}

\author{Jose Martinez-Castro}
\affiliation{Peter Gr\"unberg Institut (PGI-3), Forschungszentrum J{\"u}lich GmbH,
52428 J{\"u}lich, Germany}

\author{Markus Ternes}
\affiliation{Peter Gr\"unberg Institut (PGI-3), Forschungszentrum J{\"u}lich GmbH,
52428 J{\"u}lich, Germany}
\affiliation{Institute for Experimental Physics II B, RWTH Aachen University, 52074 Aachen, Germany}

\author{F. Stefan Tautz}
\affiliation{Peter Gr\"unberg Institut (PGI-3), Forschungszentrum J{\"u}lich GmbH,
52428 J{\"u}lich, Germany}
\affiliation{Institute for Experimental Physics IV A, RWTH Aachen University, 52074 Aachen, Germany}

\author{Felix L\"upke}
\affiliation{Peter Gr\"unberg Institut (PGI-3), Forschungszentrum J{\"u}lich GmbH,
52428 J{\"u}lich, Germany}

\author{Claus M. Schneider}
\affiliation{Peter Gr\"unberg Institut (PGI-6), Forschungszentrum J{\"u}lich GmbH,
52428 J{\"u}lich, Germany}
\affiliation{Fakult{\"a}t f{\"u}r Physik, Universit{\"a}t Duisburg-Essen, 47048 Duisburg, Germany}
\affiliation{Physics Department, University of California, Davis, CA 95616, USA}

\author{J\"urgen Henk}
\affiliation{Institut für Physik, Martin-Luther-Universität Halle-Wittenberg, 06099 Halle (Saale), Germany}

\author{Lukasz Plucinski}
\email{l.plucinski@fz-juelich.de}
\affiliation{Peter Gr\"unberg Institut (PGI-6), Forschungszentrum J{\"u}lich GmbH,
52428 J{\"u}lich, Germany}
\affiliation{Institute for Experimental Physics II B, RWTH Aachen University, 52074 Aachen, Germany}


\date{\today}

\begin{abstract}
Recent years have witnessed a steady progress towards blending 2D quantum materials into technology, with future applications often rooted in the electronic structure. Since crossings and inversions of electronic bands with different orbital characters determine intrinsic quantum transport properties, knowledge of the orbital character is essential. Here, we benchmark angle-resolved photoelectron emission spectroscopy (ARPES) as a tool to experimentally derive orbital characters. For this purpose we study the valence electronic structure of two technologically relevant quantum materials, graphene and WSe$_2$, and focus on circular dichroism that is believed to provide sensitivity to the orbital angular momentum. We analyze the contributions related to angular  atomic photoionization profiles, interatomic interference, and multiple scattering. Regimes in which initial-state properties could be disentangled from the ARPES maps are critically discussed and the potential of using circular-dichroic ARPES as a tool to investigate the spin polarization of initial bands is explored. For the purpose of generalization, results from two additional materials, GdMn$_6$Sn$_6$ and PtTe$_2$ are presented in addition. This research demonstrates rich complexity of the underlying physics of circular-dichroic ARPES, providing new insights that will shape the interpretation of both past and future circular-dichroic ARPES studies.
\end{abstract}

\maketitle

\begin{figure*}
    \centering
    \includegraphics[width=18cm]{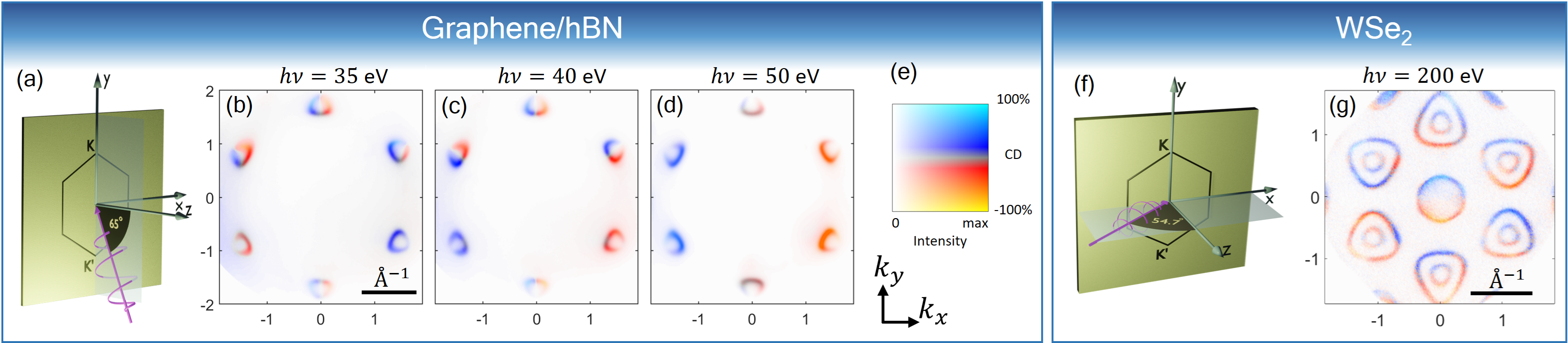}
    \caption{Experimental CD-ARPES from a monolayer of graphene on hBN (a--e) and for WSe$_2$ (f--g). (a) Experimental setup for graphene, the light is incident at $\theta_{h\nu} = 65^\circ$ off-normal. (b) -- (d) CD-ARPES maps for $h\nu = 35$, $40$, and $50$ eV, respectively, constant energy cuts are at the binding energy $E_B = 1.25$ eV. (f--g) Experimental setup and the CD-ARPES map for WSe$_2$ taken at $h\nu = 200$ eV and $\theta_{h\nu} = 54.7^\circ$. The constant energy cut in (g) is at 0.725 eV below the valence band maximum at $K/K'$ points (see Supplementary Information for details). Panels (b)--(d) and (g) are centered at normal emission and use the colorscale shown in (e). In both setups, (a) and (f), the reaction plane coincides with a mirror plane of the surface, which is reflected in the CD-ARPES maps by the sign reversal upon reflection at the mirror plane.}
    \label{fig1:CD-ARPES}
\end{figure*}

\section{Introduction and motivation}
Detailed understanding of quantum materials is a foundation upon which future information technologies will be based. Some of the key phenomena along this path are related to an intimate relation between the electronic band structure and transport properties where orbital and spin character band inversions play a key role. In particular, regions of avoided band crossings with mixed orbital angular momenta (OAM) and mixed spin characters contribute to a non-vanishing Berry curvature, an integral of which throughout the Brillouin zone (BZ) determines the conductivity within the Kubo linear response formalism.
Recently, a new field -- known as orbitronics and dealing with detection, manipulation and dynamics of OAM of electrons in solids -- has emerged and is rapidly advancing \cite{Go2021}. Since light does not directly couple to electron spin, the spin sensitivity of experimental techniques such as magneto-optical Kerr rotation and X-ray magnetic circular dichroism is a consequence of light coupling with the OAM via spin-orbit coupling (SOC). The influence of OAM-related processes goes as far as an alternative explanation \cite{Ovalle2023arXiv} of the celebrated spin-Hall experiment \cite{Kato2004}.

Circular dichroic angle-resolved photoemission (CD-ARPES) has been broadly used to reveal OAM-related effects in dispersive valence bands. An incomplete list includes studying the orbital Rashba effect \cite{Park2012}, chiral orbital angular momenta in topological surface states \cite{Park2012b}, spin textures of topological surface states \cite{Wang2011}, OAM textures in the surface states of WTe$_2$ \cite{Jiang2015}, Berry curvatures of spin-momentum-locked bands \cite{Schueler2020,Cho2018,Cho2021}, OAM textures in topological Kondo insulators \cite{Ohtsubo2022}, signatures of a spin-orbital chiral metal \cite{Mazzola2024}, time-reversal symmetry breaking in an altermagnet \cite{Fedchenko2024}, and an OAM texture of indenene \cite{Erhardt2024}. For orbitronics, proper identification of OAM of electronic states from CD-ARPES is of utmost importance, as it governs fundamental non-equilibrium effects such as orbital currents \cite{Go2021} and orbital pumping \cite{Hayashi2023}, the orbital Hall effect \cite{Go2018}, orbital relaxation \cite{Sohn2024}, laser-induced orbital magnetization \cite{Berritta2016}, and even the phenomenon of ultrafast magnetization dynamics \cite{Busch2023,Seifert2023,Nukui2024}. 

While well understood for atoms \cite{Dubs1985}, where it has been termed {\it circular dichroism in angular distribution} (CDAD), CD-ARPES, as a fundamental probe of electronic structures with its relation to the OAM and spin characters of the electronic states, remains to be thoroughly understood in solids.

Here, we analyze in detail the physics underlying the CD-ARPES process by comparing experimental maps to respective theoretical calculations. For our study we focus primarily on two technologically important materials, graphene and WSe$_2$, while spectra from a Kagome magnet GdMn$_6$Sn$_6$ and a topological metal PtTe$_2$ are also presented.

We qualitatively explain CD-ARPES maps from graphene by coherently adding multiply scattered atomic-like emissions, calculated within a real-space photoelectron diffraction formalism \cite{Abajo2001} and taking into account the two nonequivalent atomic sites. These results are compared to one-step model calculations based on the Korringa-Kohn-Rostoker formalism \cite{Gierz2011}. Subsequently, a similar analysis is discussed for WSe$_2$, a material that exhibits a dominant W 5d $Y_2^{\pm 2}$ orbital character at the $K$ and $K'$ points in the Brillouin zone. Recall that the results of one-step model calculations for graphene and WSe$_2$ were previously published in Refs.~\cite{Gierz2011,Krasovskii2022,Krueger2022,Beaulieu2020,Schusser2024}.

During the photoemission process with circularly polarized light, electrons can acquire non-zero OAM through dipole selection rules, regardless of their initial state OAM\@. This leads to OAM-dependent scattering processes, known as the Daimon effect \cite{Daimon1993}, and moreover it can lead to large CD-ARPES signals even in the absence of the initial state OAM\@. In case of graphene, we demonstrate the mutual impact of these processes on the experimental CD-ARPES maps. Concerning WSe$_2$, a further modulation of the signal related to spin-orbit coupling (SOC) is analyzed. To broaden our analysis, we also explore the CD-ARPES signals from GdMn$_6$Sn$_6$ and PtTe$_2$, highlighting several important features that showcase the versatility  of CD-ARPES across different materials.

This study aims at solving a number of puzzles that are related to the CD-ARPES technique. Moreover, it provides a foundation for deriving the initial-state OAM from CD-ARPES maps. In this way, it serves as a benchmark for interpreting past and future dichroic photoemission experiments. 


\section{Results}

Figure~\ref{fig1:CD-ARPES} shows exemplary experimental CD-ARPES maps from graphene and WSe$_2$ (additional maps are shown in the Supplementary Information). The Dirac states at the $K$ and $K'$ points of graphene are made up almost exclusively of C $2p~Y_1^0$ (that is $2p_z$) orbitals that carry zero OAM\@. Therefore, if CD-ARPES was to mirror their OAM character, the dichroic signal from the Dirac states of graphene should vanish. Instead, the maps in Fig.~\ref{fig1:CD-ARPES}(b)-(d) exhibit strong and rich dichroic signals, which vary significantly with energy. 
As discussed in the Supplementary Information, CD-ARPES maps from graphite closely resemble results obtained from graphene, suggesting that the substrate does not significantly impact the overall character of the maps.

In close vicinity of the $K$ and $K'$ points, the topmost valence bands of WSe$_2$ are primarily of W $5d~Y_2^{\pm 2}$ character, that is, they carry OAM of $\pm 2 \hbar$. If CD-ARPES was to be sensitive to the OAM, the maps should exhibit sign inversion between $K$ and $K'$ points. Instead, the map in Fig.~\ref{fig1:CD-ARPES}(g) displays a complex pattern of sign reversals, not clearly alternating between $K$ and $K'$ points.

\begin{figure*}
    \centering
    \includegraphics[width=18cm]{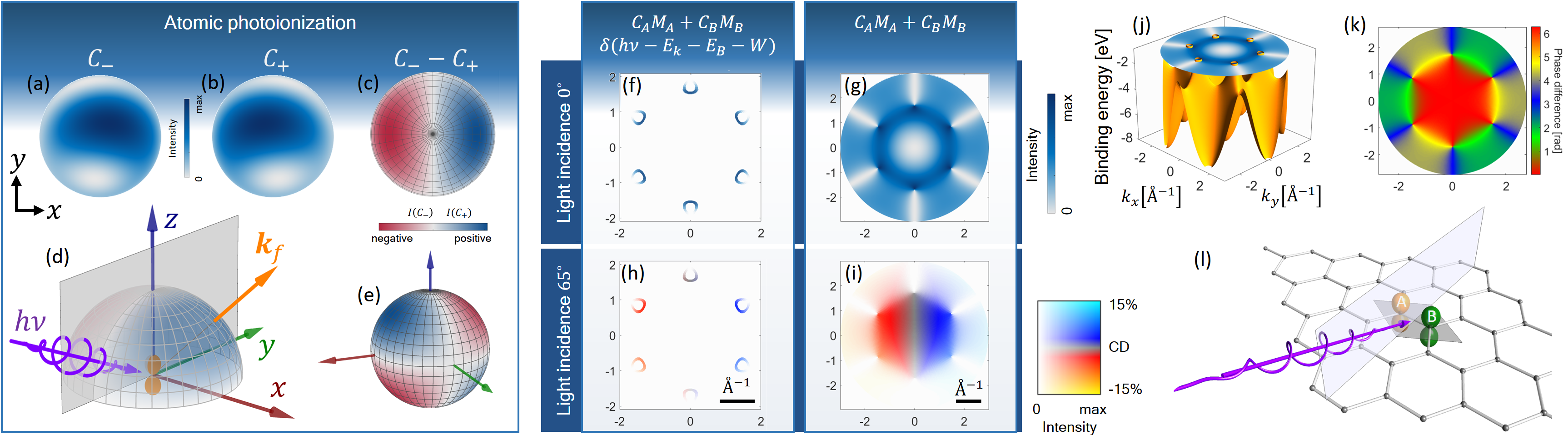}
    \caption{(a)-(e) CDAD from a C 2p $Y_1^0$ orbital at $\theta_{h\nu} = 65^\circ$ and $h\nu=40$ eV calculated using EDAC code \cite{Abajo2001}. (f)-(g) Theoretical photoemission from graphene according to Eq.~\eqref{eq:IACA} at  $\theta_{h\nu}=0^\circ$, $h\nu=40$ eV, and $E_B=1$ eV with (f) and without (g) energy conservation (confer text). The band structure is depicted in  (j). (h)-(i) As (f)-(g), but at $\theta_{h\nu}=65^\circ$. A special 2D colormap visualizes simultaneously CD and intensity. (k) Difference between complex phases of the coefficients $A(\mathbf{k_{\parallel}})$ and $B(\mathbf{k_{\parallel}})$ of graphene $\pi$ bands in the tight binding model. (l) Graphene lattice with nonequivalent sites $A$ and $B$; the grey rhombus indicates the unit cell, and the rectangle indicates the photoemission reaction plane with light incident at $\theta_{h\nu}=65^\circ$.}
    \label{fig2:EDAC_IACA}
\end{figure*}

\begin{figure*}
    \centering
    \includegraphics[width=18cm]{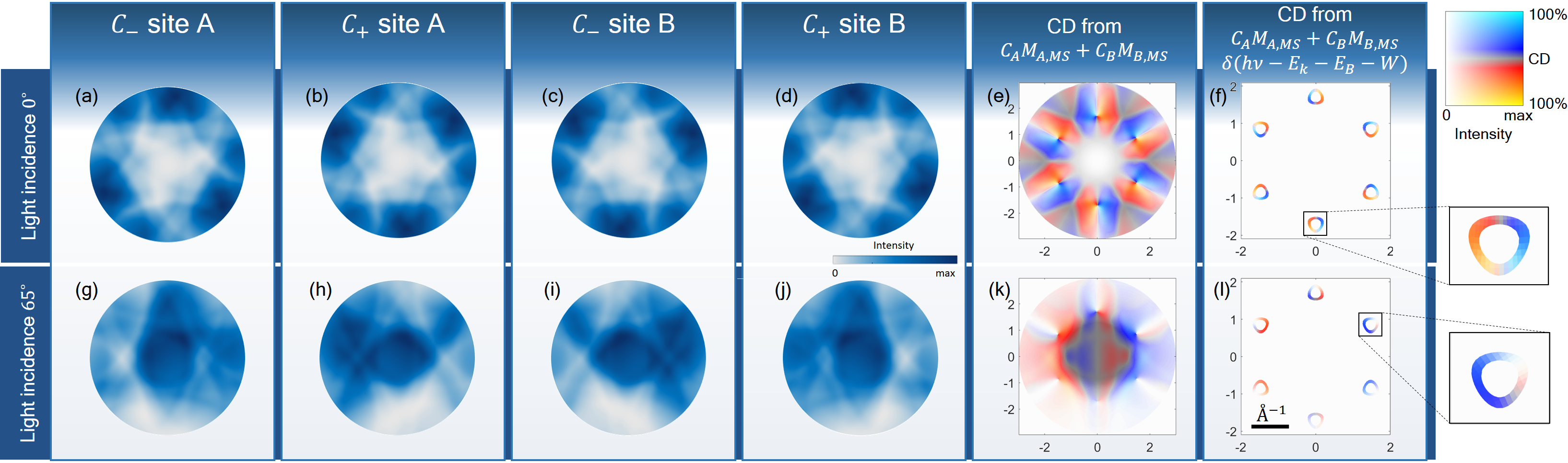}
    \caption{(a)-(d) Photoelectron diffraction patterns at $\theta_{h\nu} = 0^\circ$ and $h\nu = 40$ eV over the upper half space (compare Fig.~\ref{fig1:CD-ARPES} (d)) for the graphene sites $A$ and $B$ excited with $C_\pm$ radiation, as indicated. Patterns (a)-(d) are added coherently, taking into account amplitudes, phases, interatomic interferences, as well as the coefficients $A(\mathbf{k_{\parallel}})$ and $B(\mathbf{k_{\parallel}})$ without (e) and with (f) energy conservation $\delta(h\nu - E_k - W - E_B)$. (g)-(l) Same as (a)-(f) but at $\theta_{h\nu} = 65^\circ$. The complex phases related to (a)-(d) and (g)-(j) are shown in the Supplementary Information.} 
    \label{fig3:EDAC_MS}
\end{figure*}

In order to explain this behavior qualitatively, we employ a photoemission model based on the tight-binding formalism for the ground-state band structure and atomic-like photoionization profiles \cite{Schattke_Section_2.7}. The latter are augmented with multiple scattering in a real~space cluster \cite{Abajo2001,Krueger2011}. For a 2D solid, the model is based on the coherent addition of photoelectron emissions originating from participating $Y_l^m$ orbitals of the nonequivalent sites in the unit cell. 

In the case of graphene, there are two nonequivalent sites, labeled $A$ and $B$ in this paper. For the Dirac $\pi$ bands \cite{Reich2002} each one contributes with a single $Y_1^0$ orbital. 
In the first level of approximation, that is, the independent atomic center approximation (IACA) \cite{Krueger2022}, atomic-like photoionization profiles $M_A$ and $M_B$ due to emitters $A$ and $B$ are combined coherently. This is done taking into account parallel momentum conservation, phases of momentum-dependent complex coefficients from the tight binding model of graphene with nearest-neighbor hopping \cite{Kern2023,Reich2002}, as well as phase shifts due to positions of emitters with respect to the emission direction. The initial-state tight-binding wave function of graphene can be written as  $\psi(\mathbf{k}) = A(\mathbf{k_{\parallel}})|C_A\rangle + B(\mathbf{k_{\parallel}})|C_B\rangle$, where $|C_A\rangle$ and $|C_B\rangle$ are atomic C $2p ~Y_1^0$ wave functions centered on the two nonequivalent sites within the unit cell and  $A(\mathbf{k_{\parallel}})$ as well as $B(\mathbf{k_{\parallel}})$ are complex coefficients. The total transition matrix element then reads
\begin{equation}
\begin{aligned}
M_{fi}(\mathbf{k}_f) \propto \big( A(\mathbf{k_{\parallel}}) \cdot M_A(\mathbf k_f) +  B(\mathbf{k_{\parallel}})  \cdot M_B(\mathbf k_f) \big) \times \\  \delta(h\nu - E_k - W - E_B),
\label{eq:IACA}
   \end{aligned}
\end{equation} 
where $E_B$ is the binding energy, $W$ the work function, and $\mathbf k_f$ the wavevector of the detected far-field photoelectron.

For normal light incidence, $\theta_{h\nu}=0^\circ$, dipole selection rules require $Y_1^0 \rightarrow Y_2^{\pm 1}$ for $C_\pm$ light, and CDAD vanishes (see Supplementary Information for details). For our experimental geometry with off-normal light incidence at $\theta_{h\nu}=65^\circ$ the matrix elements $M_{A,B}$ calculated by EDAC code \cite{Abajo2001} and the resulting CDAD are shown in Fig.~\ref{fig2:EDAC_IACA} (a)-(e). The intensity profiles in Fig.~\ref{fig2:EDAC_IACA} (a)-(b) depend on the light chirality and, despite $m=0$, lead to non-vanishing CDAD patterns shown in Fig.~\ref{fig2:EDAC_IACA} (c) and (e), a result known from atomic physics \cite{Dubs1985}. 

The results for graphene according to Eq.~\ref{eq:IACA}, using atomic photoionization patterns and phases of the nearest-neighbor tight-binding model coefficients are shown in Fig.~\ref{fig2:EDAC_IACA}(f)-(g) for $\theta_{h\nu}=0^\circ$ and (h)-(i) for $\theta_{h\nu}=65^\circ$. It is instructive not only to inspect the calculated ARPES maps, Fig.~\ref{fig2:EDAC_IACA} (f) and (h), but also maps in which the $\delta(h\nu - E_k - W - E_B)$ term was not taken into account; the latter could be imagined as approximately visualizing CD-ARPES over the entire $\pi$ band, as explained in Fig.~\ref{fig2:EDAC_IACA} (j) (rigorously, this neglects that different momentum regions of the $\pi$ band are probed at different kinetic energies). Figures~\ref{fig2:EDAC_IACA} (g) and (i) exhibit dark corridors, that is, regions of vanishing or at least small intensity. These result from the blue regions in Fig.~\ref{fig2:EDAC_IACA} (k), that is, regions of phase difference of $\pi$ between $A(\mathbf{k_{\parallel}})$ and $B(\mathbf{k_{\parallel}})$ \cite{Gierz2011,Liu2011,Kern2023}; then emissions from sites $A$ and $B$ enter with opposite signs. The dark corridor emerges for emission directions where, within IACA, nonequivalent sites contribute with the same intensity, as is the case for the plane shown in Fig.~\ref{fig2:EDAC_IACA} (l), and therefore is strictly valid  within Eq.~\eqref{eq:IACA}.

Figure~\ref{fig2:EDAC_IACA} (h)-(i) shows predicted CD-ARPES according to Eq.~\eqref{eq:IACA}. Fig.~\ref{fig2:EDAC_IACA} (i) is an intensity-modulated atomic pattern of Fig.~\ref{fig2:EDAC_IACA} (c) and demonstrates that the linear combinations of atomic-like photoionization patterns, Eq.~\eqref{eq:IACA}, cannot explain the experimental patterns of Fig.~\ref{fig1:CD-ARPES} (b)-(d) that exhibit additional intra-contour CD sign reversals.

\begin{figure}
    \centering
    \includegraphics[width=8cm]{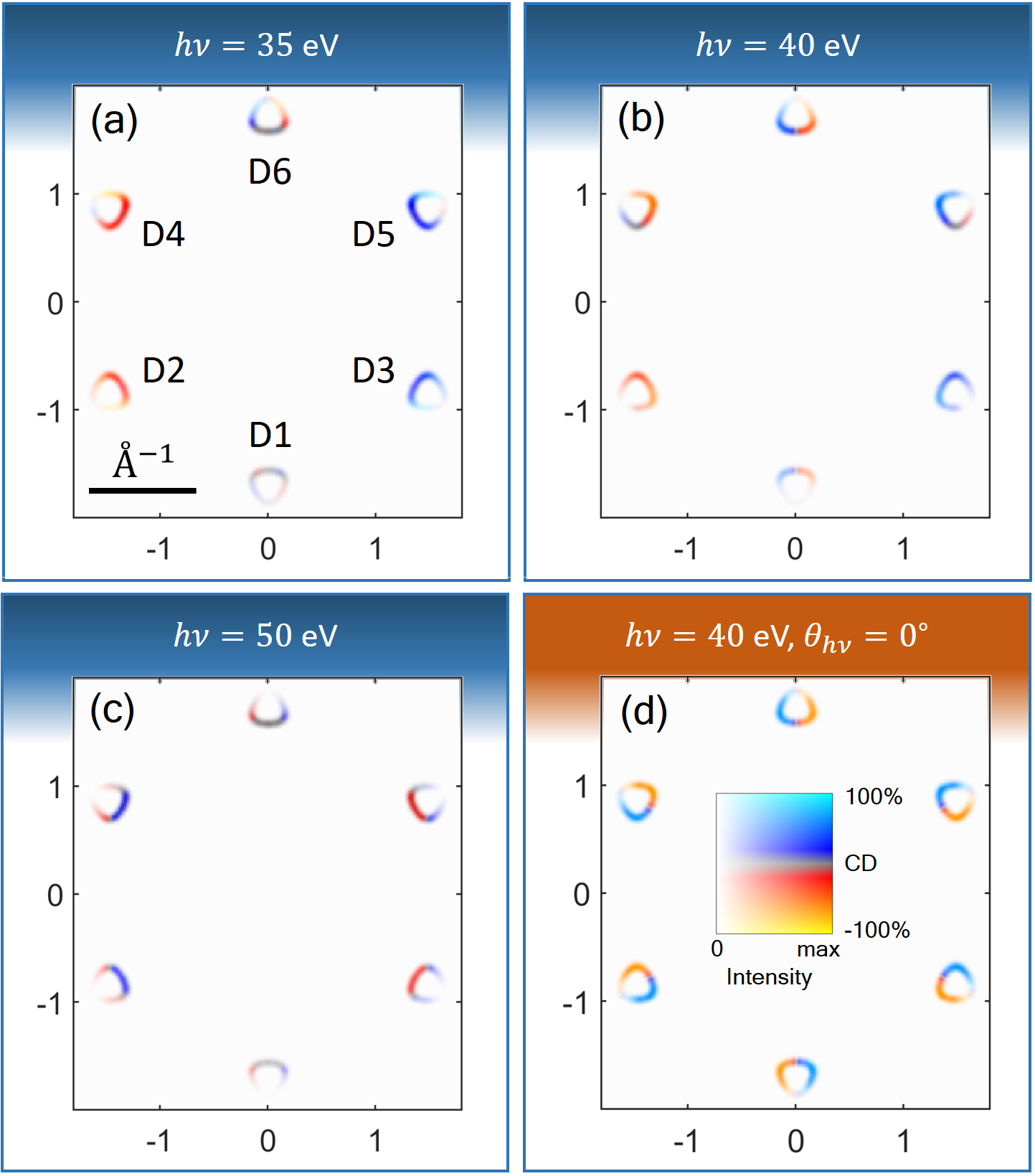}
    \caption{Theoretical photoemission calculated within the one-step model for graphene at $E_B = 0.8$ eV and $\theta_{h\nu}=65^\circ$: (a) $h\nu = 35$ eV, (b) $40$ eV, and (c) $50$ eV. (d) Same as (b) but at $\theta_{h\nu}=0^\circ$. The colormap used in all panels is shown in (d), all scales are in Å$^{-1}$.}
    \label{fig4:omni}
\end{figure}

In the following we will extend Eq.~\eqref{eq:IACA} by including photoelectron scattering. Conceptually, this is done in a straightforward way by replacing atomic-like matrix elements $M(\mathbf k_f)$ by their multiple scattered (MS) counterparts $M_{MS}(\mathbf k_f)$. In this approach a cluster of atoms around the emitter site is considered, the scattered wave originating from the emitter site is calculated and used as a final state in $M_{MS}(\mathbf k_f)$  \cite{Abajo2001,Krueger2011,Matsui2018}. Both elastic and inelastic processes can be taken into account, the latter ones by including a finite inelastic mean free path (IMFP). This type of calculation is performed within the photoelectron diffraction formalism \cite{Sebilleau2006,Matsushita2010,Krueger2022}, here using EDAC \cite{Abajo2001}. The results are shown in Fig.~\ref{fig3:EDAC_MS}, with panels (a)-(f) related to $\theta_{h\nu} = 0^\circ$ and (g)-(l) to $\theta_{h\nu} = 65^\circ$. 

\begin{figure*}
    \centering
    \includegraphics[width=16cm]{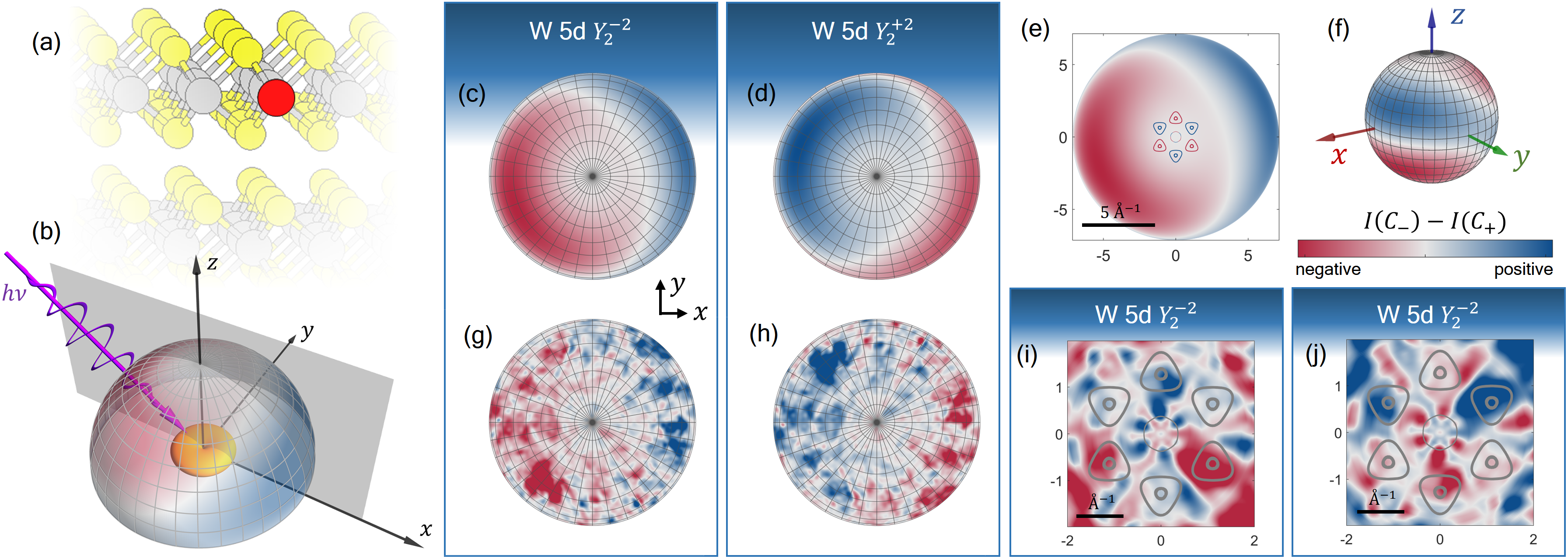}
    \caption{Photoelectron diffraction from W 5d $Y_2^{\pm 2}$ orbitals in WSe$_2$ at $h\nu=200$ eV calculated using EDAC \cite{Abajo2001}. (a) Outermost layers of WSe$_2$ with an emitter W atom indicated in red. (b) Experimental geometry with the reaction plane (grey rectangle) and the $10^\circ$ step grid in polar and azimuthal angles; the orange contour represents a $Y_2^{\pm 2}$ orbital. (c)-(d) CDAD signal from W $5d$ $Y_2^{-2}$ and $Y_2^2$ orbitals, respectively. (e) Same as (c), but converted to the momentum scale and with overlaid theoretical contours at $E_{VBM}-0.5$ eV for bulk WSe$_2$. The contours at the three $K$ points are related to the $Y_2^{2}$ map, while those at the $K'$ points would be related to the $Y_2^{-2}$ map. Such alternation of the CD sign would provide OAM sensitivity under IACA (our assignment of $K/K'$ is arbitrary). (f) Same as (c), but for emission angles over the full sphere. (g)-(h) Same as (c)-(d), but with multiple scattering included. (i)-(j) Same as (g)-(h), but converted to the momentum scale and with theoretical contours as in (e). Maps for other kinetic energies are shown in the Supplementary Information.}
    \label{fig5:WSe2_atomic}
\end{figure*}

Let us first focus on $\theta_{h\nu} = 0^\circ$. Panels (a)-(d) show $|M_{A,MS}(\mathbf k_f)|^2$ and $|M_{B,MS}(\mathbf k_f)|^2$ for $C_\pm$ excitations, over the half space above the surface (see Fig.~\ref{fig2:EDAC_IACA}(d)). These patterns differ significantly from the dipole-allowed atomic $|Y_2^{\pm 1}|^2$ patterns (see Supplementary Information), and their symmetries reflect the trigonal environment of nearest neighbors, together with the intensity modulations due to the Daimon effect \cite{Daimon1993}. The patterns are chiral, for each site the chirality is reversed upon reversing the light helicity. Furthermore, patterns are connected by various mirror reflections related to the fact that a mirror reflection through any plane parallel to $z$ reverses the light helicity, and to the glide reflections that connect local environments of sites $A$ and $B$ (lattice translations do not influence far field intensity patterns), for example (a) becomes (b) upon $\mathcal M_y$ and (a) becomes (d) upon $\mathcal M_x$.

The results of coherently combining the patterns of Fig.~\ref{fig3:EDAC_MS}(a)-(d) are shown in (e)-(f), without and with taking into account the energy conservation $\delta(h\nu - E_k - W - E_B)$ (similarly to Fig.~\ref{fig2:EDAC_IACA}(f)-(g)). Since at $\theta_{h\nu}=0^\circ$, atomic-like CDAD vanishes, the CD-ARPES pattern in Fig.~\ref{fig3:EDAC_MS}(f) stems exclusively from final-state scattering, that is a variant of the Daimon effect \cite{Daimon1993} for valence electrons. The magnitude of the CD asymmetry
\begin{align}
\frac{|M_{fi,MS,C_-}|^2 - |M_{fi,MS,C_+}|^2}{|M_{fi,MS,C_-}|^2 + |M_{fi,MS,C_+}|^2} 
\end{align}
reaches $\approx 80\%$.

Figure~\ref{fig3:EDAC_MS} (g)-(j) shows angular intensity patterns for $\theta_{h\nu}=65^\circ$. Compared to panels (a)-(d) some symmetries are missing, with the remaining $\mathcal M_x$ mirror operation that connects (h) with (j) and (h) with (i), as expected from the reversal of light helicity together with glide mirror transformation swapping sites $A$ and $B$. Importantly, the patterns differ qualitatively from the atomic-like patterns of Fig.~\ref{fig2:EDAC_IACA}(a)-(b). The CD-ARPES pattern in Fig.~\ref{fig3:EDAC_MS}(k) exhibits similarities to the IACA pattern in Fig.~\ref{fig2:EDAC_IACA}(c), but only concerning the intensity, the CD pattern is qualitatively different. The magnitude of the CD again reaches about $80 \%$. The dark corridor is strict (zero signal) neither in Figure~\ref{fig3:EDAC_MS}(e) nor in (k), in agreement with one-step model calculations \cite{Gierz2011}. Along the $k_x = 0$ momentum trajectory the intensity reaches $\approx 5\%$ of the maximum intensity.

\begin{figure}[!ht]
    \centering
    \includegraphics[width=8cm]{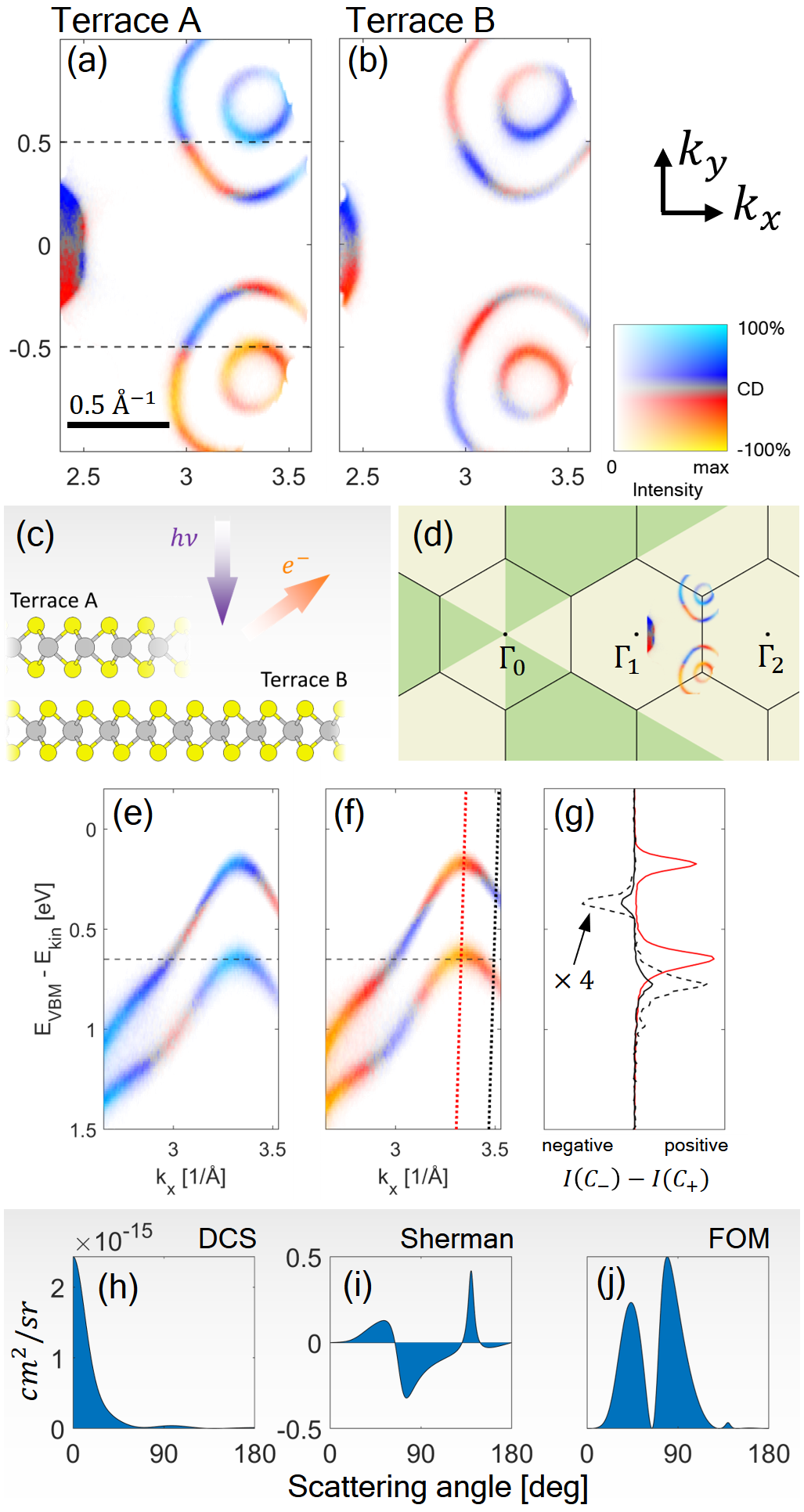}
    \caption{Photoemission from WSe$_{2}$. (a)-(b) Constant-energy cuts at $E_{VBM}-E_{kin} = 0.64$ eV for terraces $A$ and $B$ at $h\nu = 60$ eV and $\theta_{h\nu} = 0^\circ$ as indicated in (c). $E_{VBM}$ refers to the valence band maximum at the $K/K'$ points; the assignment of terraces in (c) is arbitrary. (d) Position of (a) in momentum space. (e)-(f) Energy-momentum cuts along the dashed lines in (a). The horizontal dashed line indicates the energy in (a) and (b). (g) Red and black solid curves show the intensity difference $I(C_-) - I(C_+)$ along red and black dotted lines in (f). The dashed black line is multiplied by a factor of $4$. (h) Differential cross section (DCS) for scattering of free electrons at W-atoms, calculated using ELSEPA \cite{Salvat2021} with Sherman function $S$ (i) and figure-of-merit DCS$\times S^2$ (j) \cite{Jost1981}.}
    \label{fig6:WSe2_normal_incidence}
\end{figure}

\begin{figure*}[!ht]
    \centering
    \includegraphics[width=18cm]{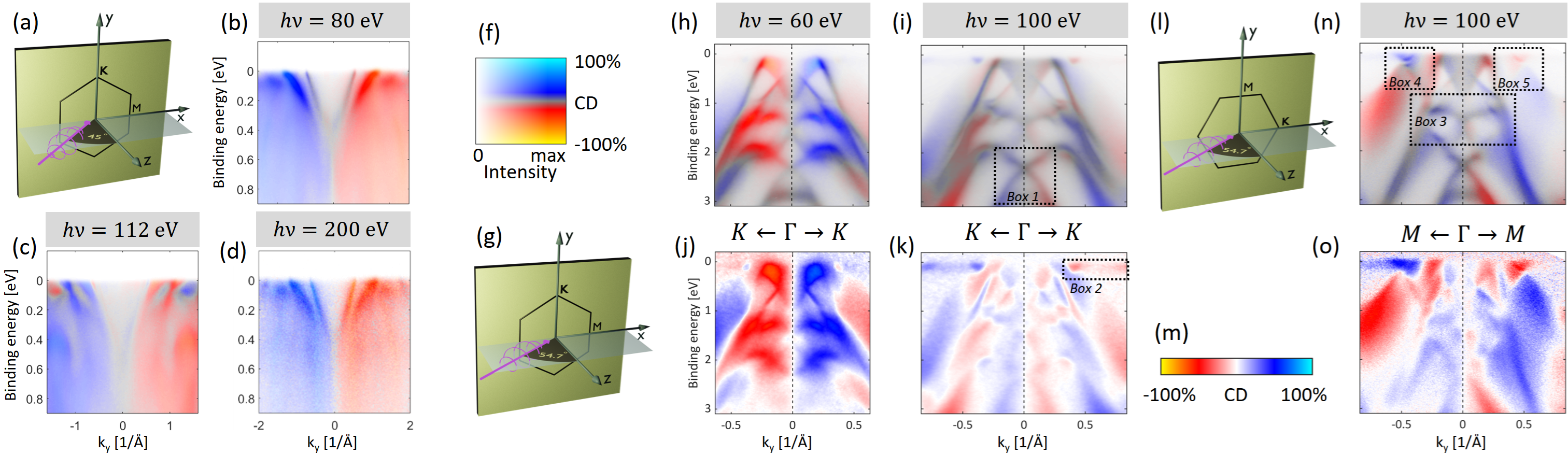}
    \caption{(a)-(d) Experimental geometry and CD-ARPES energy-momentum maps from GdMn$_6$Sn$_6$ along $\Gamma K$ reciprocal direction at $h\nu = 80$ eV (b), $h\nu = 112$ eV (c), and $h\nu = 200$ eV (d). Panels are plotted using the 2D colormap shown in (f). (g)-(o) Experimental geometries and CD-ARPES energy momentum maps from PtTe$_2$. Experimental geometry in (g) refers to panels (g)-(j), while the geometry in (l) to panels in (l)-(m). (h) and (j) were taken at $h\nu = 60$ eV, while (j), (k), (n), and (o) were taken at $h\nu = 100$ eV. Upper panels (h), (i), and (n) are plotted using the 2D colormap shown in (f), while lower panels (j), (k), and (o) show CD magnitude according to the colormap shown in (m). Spectra taken at $\approx 40$K. Features indicated by $boxes$ 1 to 4 in (i), (k), and (n) are discussed in the text.}
    \label{fig7:NEW_FIGURE}
\end{figure*}

In order to validate the findings of our real-space scattering model, we compare the experimental CD-ARPES maps taken at $h\nu=$ 35, 40, and 50 eV shown in Fig.~\ref{fig1:CD-ARPES}(b)-(d) with respective results from one-step model photoemission calculations shown in Fig.~\ref{fig4:omni}(a)-(c), performed with the previously used setup \cite{Gierz2011}. In this energy range, the sign of the experimental CD behaves differently for different Dirac contours, which we label D1-D6 (cf.\ Fig.~\ref{fig4:omni}(a)). The experimental behavior is partly reproduced in the one-step model calculations, for example the reversal of the CD sign between $h\nu=$ 35 and 50 eV for contours D2 and D3. Also the sign change of the intra-contour CD is reproduced for D4-D6 (Fig.~\ref{fig1:CD-ARPES}(c) and Fig.~\ref{fig4:omni}(b)), as is the weakening of intra-contour sign flip for D1 and D6 in Fig.~\ref{fig1:CD-ARPES}(d) and Fig.~\ref{fig4:omni}(c). On the other hand, the sign of contours D2-D3 is not correctly reproduced (panels Fig.~\ref{fig1:CD-ARPES}(c) and Fig.~\ref{fig4:omni}(b)).

So far we discussed the physics of non-vanishing CD-ARPES in graphene, a material in which initial states carry negligible OAM and spin polarization. In the following we will extend the discussion to WSe$_2$, in which the inital states carry orbital and spin polarization. In WSe$_2$ near the  $K/K'$ points, the states at the valence band maximum are primarily of W 5d $Y_2^{\pm 2}$ character with an additional spin-polarized splitting \cite{Riley2014}. As discussed in the Supplementary Information, the admixture of other orbitals near $K/K'$ is not negligible, nevertheless in the following we will focus on dominant contributions by W 5d $Y_2^{\pm 2}$.

Figure~\ref{fig5:WSe2_atomic} shows theoretical photoelectron diffraction results at $h\nu = 200$ eV from a W 5d $Y_2^{\pm 2}$ orbital in the surface layer of bulk WSe$_2$ (Fig.~\ref{fig5:WSe2_atomic}(a); the coordinate system is defined in Fig.~\ref{fig5:WSe2_atomic}(b)). The atomic-like CDAD profiles from $Y_2^{2}$ and $Y_2^{-2}$ exhibit sign reversals, as shown in Fig.~\ref{fig5:WSe2_atomic} (c) and (d), respectively. This indicates that already within IACA different CD signs can be expected for different emission angles \cite{Sagehashi2023}, and there may exist regions where the CDAD vanishes. On the other hand, the positions of these sign changes vary slowly with emission angle, and, as shown in Fig.~\ref{fig5:WSe2_atomic} (e), the entire first BZ encloses the same sign of atomic CDAD near normal emission at $h\nu=200$ eV. This demonstrates that rapid CD sign changes (see Fig.~\ref{fig1:CD-ARPES} (g)) in the vicinity of a $K$ or $K'$ point are unlikely to originate from atomic-like CDAD patterns, even if one considers coherent summation of such patterns from all sites in the spirit of Eq.~\eqref{eq:IACA}. Figure~\ref{fig5:WSe2_atomic} (f) shows the pattern of emission over the full sphere, which, when angle-integrated, will yield a net CD signal. This reflects a net sensitivity of angle-integrating methods based on core-level absorption to the OAM\@. However, in such methods typically only the $l+1$ channel is significant (e.g. $2p$ to $3d$ channel in transition metals), and the three-step model description is sufficient (we discuss this further in the Supplementary Information). 

Figure~\ref{fig5:WSe2_atomic}~(g-h) shows the photoelectron diffraction maps for $Y_2^{\pm 2}$ orbitals with multiple scattering included. Figure~\ref{fig5:WSe2_atomic}~(i-j) shows portions of the same maps, converted to the momentum scale and with overlaid theoretical contours. These results qualitatively explain rapid variations of the CD-ARPES signal from WSe$_2$ as being due to multiple scattering.

With the quantization axis along the surface normal, the CDAD vanishes for orbitals with $m=0$ for normal light incidence, potentially simplifying the interpretation of CD-ARPES maps. Figure~\ref{fig6:WSe2_normal_incidence}~(a)-(b) shows CD-ARPES results from WSe$_2$ at normal light incidence and $h\nu = 60$ eV from the terraces $A$ and $B$, as shown in (c). In these experiments, the angle between the incoming light and normal emission is fixed at $54.7^\circ$, therefore the maps in Fig.~\ref{fig6:WSe2_normal_incidence}~(a)-(b) are probing a momentum section between the second and third BZ, as shown in (d). Since the $K$ and $K'$ points are swapped at subsequent terraces \cite{Gehlmann2016, Riley2014}, any swapping of the CD sign between $K$ and $K'$ in maps in Fig.~\ref{fig6:WSe2_normal_incidence}~(a)-(b) can be potentially related to the initial state OAM. Indeed such CD sign reversals are present, however, only along sections of the $K$ and $K'$ contours. Inner contours exhibit constant CD sign in  Fig.~\ref{fig6:WSe2_normal_incidence}~(a) while the intra-contour sign reversal is present in (b). The difference between Fig.~\ref{fig6:WSe2_normal_incidence} (a) and (b) is due to the final state scattering combined with the atomic final state profile because terraces $A$ and $B$ have reversed polarities along the $x$-axis, which causes a difference in scattering. This effectively divides the CD-ARPES maps into non-equivalent regions, as indicated by beige and green segments in Fig.~\ref{fig6:WSe2_normal_incidence}~(d). 

Figure~\ref{fig6:WSe2_normal_incidence}~(e)-(f) shows energy-momentum cuts along the dashed lines in (a) that are in the vicinities of $K/K'$ points. At some momenta the CD sign of the two split bands is opposite, as shown by black curves in Fig.~\ref{fig6:WSe2_normal_incidence}~(g). Since within a layer these states carry the same OAM but opposite spin polarization \cite{Riley2014,Gehlmann2016} (see the Supplementary Information for details), this finding can be related to spin-orbit scattering in the final state \cite{Gregory1974,Jost1981}. To estimate this effect, it is useful to study differential cross section and the Sherman function of W atoms, both shown in Fig.~\ref{fig6:WSe2_normal_incidence}(h)-(i) together with their figure-of-merit (FOM) in (j) \cite{Kessler1985,Jost1981}. Indeed, at $E_{kin} = 60$ eV, a W atom is expected to produce significant spin polarization at scattering angles between 30$^\circ$ and 100$^\circ$. When considering various W and Se emitters and nearest-neighbor and next-nearest-neighbor scatterers, this may lead to the spin-dependent inversions of the CD sign shown in Fig.~\ref{fig6:WSe2_normal_incidence}~(f)-(g).


Figure \ref{fig7:NEW_FIGURE} shows CD-ARPES energy-momentum maps for two additional materials, a Kagome magnet GdMn$_6$Sn$_6$, where electron correlations play a role \cite{Liu2021}, and a topological metal PtTe$_2$ that can be described by one-electron physics \cite{Masilamani2024,Yan2017,Bahramy2017}.

GdMn$_6$Sn$_6$ data has been measured in the experimental geometry indicated in Fig.~\ref{fig7:NEW_FIGURE}~(a), where the reaction plane overlaps with one of the sample mirror planes. Through the axial vector mirror reflection rules, this implies that CD-ARPES maps are odd in $k_y$, as indeed exemplified in Fig.~\ref{fig7:NEW_FIGURE}~(b)-(d). Importantly, all three panels exhibit dominating CD sign at negative and positive $k_y$, with local CD sign reversals only observed in (c) and only for the quasiparticle bands near the Fermi level. For $E_B > 0.6$ eV correlated electrons exhibit a simple CD pattern, with no CD sign reversals on either side of $k_y=0$. This suggests that the observed behavior might be primarily due to the IMFP-derived CD, as discussed by Moser \cite{Moser2017}.

The CD-ARPES maps from PtTe$_2$ were measured in two geometries. In the geometry of Fig. \ref{fig7:NEW_FIGURE} (g) the reaction plane coincides with the mirror plane of the sample, and therefore the CD maps in (h)-(k) are odd in $k_y$. We present the maps in two ways; panels in Fig. \ref{fig7:NEW_FIGURE} (h)-(i) are plotted using the 2D colormap of (g), while in panels (j)-(k) only the absolute CD signal is plotted, according to the colormap in (m). PtTe$_2$ exhibits a surface Dirac cone centered at $\overline \Gamma$ at $E_B$ between 2 and 3 eV \cite{Masilamani2024,Yan2017,Bahramy2017}, as indicated in $Box~1$ in Fig. \ref{fig7:NEW_FIGURE} (i). This Dirac cone is expected to exhibit spin-momentum locking, which leads to the two branches having opposite spin polarization. It has been shown that for topological insulators CD-ARPES can potentially be a probe of this spin-momentum locking \cite{Barriga2014, Wang2011}, however this behavior is not observed in our experiments, where, e.g. for positive $k_y$ both upper and lower branches of the Dirac cone exhibit the same CD sign.

By comparing the panels (h)-(i) with (j)-(k) we demonstrate that non-vanishing CD signal is present also in the regions of the expected projected band structure gaps. As exemplary indicated in Fig. \ref{fig7:NEW_FIGURE} (k) by $Box~2$, a resulting flat CD-ARPES band appears at the Fermi level, where inelastic scattering is expected to play a minor role. 

In the geometry of Fig. \ref{fig7:NEW_FIGURE} (l) mirror planes of the crystal are broken by the experimental geometry ($\Gamma -K$ is not a mirror plane of PtTe$_2$) and therefore CD-ARPES maps in (n)-(o) are not symmetric in $k_y$, as indicated for example by $Box~3$ in (n) where both the CD sign and the energy positions of bands are asymmetric for $\pm k_y$. These asymmetries stem primarily from interatomic interferences \cite{Heider2023} and asymmetric multiple scattering. In addition the trigonal symmetry of the bulk band structure, in connection with approximate $k_\bot$ sensitivity of ARPES, may also play a role.

Projected band structure is made from overlapping allowed continuous regions in energy-momentum maps that often originate primarily from one type of orbital \cite{Shockley1939}, therefore such different regions can be expected to have different responses in CD. $Boxes$ 4 and 5 in  Fig. \ref{fig7:NEW_FIGURE} (n) illustrate the regions where such sensitivity appears to take place, with the hole pockets that host surfaces state \cite{Yan2017} having opposite CD sign to the surrounding projected bands (the same applies to the tiny hole pockets in (i)). However, it remains to be established under which conditions this contrast can be considered as a faithful representation of the initial state band character.


\section{Discussion \label{sec:Discussion}}

Our results for graphene show that through the Daimon effect \cite{Daimon1993} strong CD-ARPES is present even in the absence of SOC\@. Therefore, in materials in which SOC is relevant, isolating the initial state OAM from CD-ARPES maps is a hard problem. For a material of the WSe$_2$ family, where bands with $m = \pm 2$ alternate between the $K$ and $K'$ points, the breaking of the intra-layer inversion symmetry that splits the $m=\pm 2$ derived bands is also responsible for asymmetries in the CD photoelectron diffraction maps. These asymmetries originate not only from the final-state scattering, but also from interatomic interferences, with the phase shift $\exp( i \mathbf{k}_f \cdot (\mathbf{r}_A-\mathbf{r}_B))$ already present in the free-electron final state approximation, and are relevant for systems with band characters mixed between any different sites $A$ and $B$ within the unit cell \cite{Heider2023}. 
Therefore, no linear combination of dichroic signals taken at different geometries 
can {\it a priori} isolate contributions from the initial state OAM from those related to interatomic interference and mutiple scattering. 


Without prior knowledge of the material, it is unlikely to deduce that the patterns shown in Fig.~\ref{fig1:CD-ARPES}(b)-(d) originate from $m=0$ orbitals. These patterns reflect the complexity of the photoemission process, for which not only multiple scattering, but also accurate modeling of the surface barrier and the optical potential \cite{Henk1993,Henk1999} can be critical. Further progress can be made by augmenting present state-of-the-art one-step model computer codes with a full-potential approach \cite{Huhne1998} instead of relying on the muffin-tin approximation.


Since CD-ARPES is a technique at least an order of magnitude faster compared to spin-polarized ARPES, for it does not require a special detector system, probing spin polarization using CD-ARPES has been long sought \cite{Wang2011}. Figure~\ref{fig6:WSe2_normal_incidence}~(f)-(g) suggests how spin sensitivity of CD-ARPES can be achieved through spin-orbit scattering in the final state at tungsten (high $Z$) atoms. A possible scenario is that CDAD and the Daimon effect vanish at a certain $\mathbf{k}_f$, allowing SOC scattering to be observed. Indeed the CD sign reversal in Ref.~\ref{fig6:WSe2_normal_incidence}~(g) appears in a region in which the net CD signal is small. However, to confirm these conjectures further numerical calculations are necessary in order to eliminate effects due to small differences in orbital contributions between the two split bands near the $K/K'$ points. 

In Fig. \ref{fig7:NEW_FIGURE} we demonstrate that the complexity and diverse physics of CD-ARPES is not limited to particular materials and opens up new possibilities to study electronic structure of quantum materials. Further extension of our work may be in spin-dependent transmission of electrons through few atomic layers, in understanding OAM and spin polarization of hot electrons after an excitation by ultrashort laser pulse \cite{Busch2023}, and in designing schemes to generate OAM electron beams through photoemission \cite{Takahashi2015}.

\section{Acknowledgements}

We would like to thank S. Nemsak for fruitful discussions, F. J. Garcia de Abajo for making the EDAC code available, and C. Jozwiak, A. Bostwick, and E. Rotenberg for the assistance at the MAESTRO beamline at ALS.

H.B. was supported by the DFG via the Project PL 712/5-1. X. H. was supported by Deutsche Forschungsgemeinschaft (DFG, German Research Foundation) under Germany’s Excellence Strategy - Cluster of Excellence Matter and Light for Quantum Computing (ML4Q) EXC 2004/1 - 390534769. M. Q. was supported by the Federal Ministry of Education and Research of Germany in the framework of the Palestinian-German Science Bridge (BMBF grant number 01DH16027).

Y.M. acknowledges support  by the EIC Pathfinder OPEN grant 101129641 ``OBELIX'' and by the DFG $-$ TRR 288 $-$ 422213477 (project B06). K.J., T.W., J.M.C. and F.L. acknowledge funding from the Deutsche Forschungsgemeinschaft (DFG, German Research Foundation) within the Priority Programme SPP 2244 (project nos. 443416235 and 443405092). J.M.C. acknowledges funding from the Alexander von Humboldt Foundation. F.L. acknowledges funding from the Emmy Noether Programme of the DFG (project no. 511561801) and Germany’s Excellence Strategy - Cluster of Excellence Matter and Light for Quantum Computing (ML4Q) through an Independence Grant. F.L. and F.S.T acknowledge funding from the Bavarian Ministry of Economic Affairs, Regional Development and Energy within Bavaria’s High-Tech Agenda Project ''Bausteine f\"ur das Quantencomputing auf Basis topologischer Materialien mit experimentellen und theoretischen Ans\"atzen''.

This publication was partially developed under the provision of the Polish Ministry and Higher Education project "Support for research and development with the use of research infra-structure of the National Synchrotron Radiation Centre SOLARIS” under contract no 1/SOL/2021/2. We acknowledge SOLARIS Centre for the access to the PHELIX beamline, where part of the measurements were performed.

\section{Methods}
\subsection{ARPES based on photoelectron diffraction}

With the tight-binding (LCAO) initial wave function $\psi_i = \sum_{l,m,\mathbf{r}_j} C_{l,m,\mathbf{r}_j} ~\phi_{l,m,\mathbf{r}_j}$ the dipole matrix element for optical transition reads
\begin{equation}
\begin{aligned}
M_{fi}(\mathbf{k}_f) \propto \langle \psi_f | \mathbf{A}\cdot \mathbf{p} | \sum_{l,m,\mathbf{r}_j} C_{l,m,\mathbf{r}_j} ~\phi_{l,m,\mathbf{r}_j} \rangle \\ \times~ \delta(h\nu - E_k - W - E_B), 
\label{eq:IACA_full_1}
   \end{aligned}
\end{equation} 
where $\phi$ are atomic wave functions  at all sites $\mathbf{r}_j$ and for all participating quantum numbers $l,m$; $C$ are complex coefficients. The matrix element may be rewritten as 
\begin{equation}
\begin{aligned}
M_{fi}(\mathbf{k}_f) \propto  \sum_{l,m,\mathbf{r}_i} C_{l,m,\mathbf{r}_j} \cdot \langle \psi_f | \mathbf{A}\cdot \mathbf{p} |   ~\phi_{l,m,\mathbf{r}_j} \rangle \\ \times~ \delta(h\nu - E_k - W - E_B). 
\label{eq:IACA_full_2}
   \end{aligned}
\end{equation} 
Let us define $M_{l,m,\mathbf{r}_i}(\mathbf{k}_f) = \langle \psi_f | \mathbf{A}\cdot \mathbf{p} | ~\phi_{l,m,\mathbf{r}} \rangle$. In photoectron diffraction codes the dipole operator $\mathbf{A}\cdot \mathbf{p}$ is often replaced by the length form $\boldsymbol{\varepsilon}\cdot \mathbf{r}$ \cite{Abajo2001}, which is appropriate because of the localized nature of $\phi$. The key difficulty in evaluating $M_{l,m,\mathbf{r}_i}(\mathbf{k}_f)$ is in finding the $\psi_f$. An advantage of photoelectron diffraction codes is that $\psi_f$ is effectively spatially limited to the region surrounding $\phi_{l,m,\mathbf{r}}$, exploiting the inelastic mean free path and a decay of spherical waves with the distance from the emitter  $\phi_{l,m,\mathbf{r}_j}$.

The final state $\psi_f$ is the time-reversed LEED state, therefore it is a sum of outgoing plane waves $e^{i\mathbf{k}_f\cdot \mathbf{r}}$ and spherical waves incoming into each site \cite{Krueger2018,Liebsch1976}. A LEED state is a sum of incoming plane wave $e^{-i\mathbf{k}_f\cdot \mathbf{r}}$ and spherical waves outgoing from each site. With muffin-tin potentials, the IACA approximation corresponds to neglecting a contribution of spherical waves, either emitted or reflected from neighboring sites, to the final state at the considered site, and as a consequence also neglecting any multiple scattering.

In case of the EDAC code \cite{Abajo2001} it is preferred that $\phi_{l,m,\mathbf{r}}$ is contained within its muffin-tin (MT) sphere. According to our WIEN2k calculations \cite{Blaha2020} for nearly touching MT spheres of C $2p$ in graphene, only 39\% of the charge is within a MT sphere. In WSe$_2$ $\approx 72$\% of the charge of W $5d$ and 59\% of the charge of Se $4p$ orbitals is within their MTs. This charge leakage out of MT spheres and overlap with neighboring MTs is affecting the evaluation of spherical wave function due to the emitter \cite{Abajo2001, Liebsch1976}; however this might not be critical since in general intensities of PED patterns are known to reflect atomic photoionization profiles, and large scattering contributions observed in CD-ARPES occur because they reflect intensity differences (between patterns taken with $C_\pm$ light). At least for the graphene $\pi$ bands, the leaking effect might be somewhat minimized, because the C $2p~Y_1^0$ orbitals primarily extend out of plane, where there are no nearest neighbor MTs. To mitigate some of these issues, Kr\"uger \cite{Krueger2018} used the acceleration form of the dipole matrix element. 



Furthermore, one may consider that the photoemission process creates a localized Coulomb hole within the electron gas \cite{Grobman1978,Moser2023}, and that the final state is represented by a partial wave expansion of the free (scattering) states of this potential. However, such effects can often be neglected, especially in metals where the photohole is screened. EDAC code \cite{Abajo2001} accounts for this effect by adding a screened photohole to the MT potential of the emitter. 

Since for a periodic solid the photoemission signal derived from Eq.~\eqref{eq:IACA_full_2} has a form of a Bloch sum, the parallel-momentum conservation $\mathbf{k}_{f\parallel} = \mathbf{k}_{i\parallel}$ is obeyed and the sum in Eq.~\eqref{eq:IACA_full_2} can be limited to orbitals within a unit cell only, when using momentum-dependent coefficients $C_{l,m,\mathbf{r}_j}(\mathbf{k}_{i\parallel})$. A more detailed discussion on this and related issues in the context of modern PED is presented for instance in Refs.~\cite{Matsui2018,Krueger2011,Schattke_Section_2.7}.

Because of the coherent propagation, our approach can be considered as a one-step model. Key approximations are related to the LCAO wave function, muffin-tin potentials (in contrast to more precise full-potential methods), and to the charge leaking out of the MT spheres. Furthermore, in EDAC the surface barrier is approximated by a potential step, while more accurate modeling is known to affect the results considerably \cite{Henk1993,Henk1999}.

\subsection{Tight-binding model for the $\pi$ bands of graphene}
Graphene is a single honeycomb layer of carbon atoms. In its unit cell, there are two carbon atoms $C_A$ and $C_B$ separated by $a_{CC} \approx 1.42$ \AA. Bands that form Dirac cones of graphene are primarily made ($>97$\% contribution, according to our WIEN2k \cite{Blaha2020} calculations) of  C $2p_z$ orbitals. The nearest-neighbor hopping Hamiltonian for these orbitals has a well-known form 
\begin{equation}\label{eq1}
\begin{aligned}
H(\mathbf k) = t \sum_i (\sigma_x \cos{\mathbf k \cdot \mathbf a_i} - \sigma_y \sin{\mathbf k \cdot \mathbf a_i} )
   \end{aligned}
\end{equation} 
where $a_i$ are the three vectors connecting nearest neighbors, and $\sigma_{x,y}$ are Pauli matrices. With $t = -2.7$ eV this Hamiltonian approximates well the graphene band structure at energies close to the Dirac points, as a comparison to the DFT calculations shows \cite{Reich2002}.

\subsection{IACA and MS photoemission models for graphene}

Let us start with atomic-like photoionizatition profiles for C atoms in graphene. With the light incidence along the surface normal, dipole selection rules require the $Y_1^0 \rightarrow Y_2^{\pm 1}$ for $C_\pm$ light. Hence, circular dichroism in angular distrubution (CDAD) vanishes (see Supplementary Information for details). The situation is different if the light incidence is not aligned with the orbital quantization axis, as shown in Figure~\ref{fig2:EDAC_IACA} (a)-(e). Treatment of this case requires either a decomposition of the orbital along the quantization axis defined by the light incidence \cite{Dubs1985PRL} or decomposition of light polarization vector along the $z$-axis \cite{Moser2017}, with both $l\pm 1$ channels being allowed and interfering. The intensity profile depends on the radial integrals and phase shifts between the $l-1$ and $l+1$ scattering states, both depending on the kinetic energy \cite{Goldberg1981}. The intensity profiles in Fig.~\ref{fig2:EDAC_IACA}~(a)-(b) depend on the light chirality and produce the CDAD pattern shown in Fig.~\ref{fig2:EDAC_IACA}~(c), a result known from atomic physics \cite{Dubs1985}. 



The initial state tight-binding wave function of graphene can be written as  $\psi(\mathbf{k}) = A(\mathbf{k_{\parallel}})|C_A\rangle + B(\mathbf{k_{\parallel}})|C_B\rangle$, where $|C_A\rangle$ and $|C_B\rangle$ are atomic-like C $2p ~Y_1^0$ wave functions centered on the two nonequivalent sites within the unit cell, with complex coefficients $A(\mathbf{k_{\parallel}})$ and $B(\mathbf{k_{\parallel}})$. The transition matrix element can be written as
\begin{equation}
\begin{aligned}
M_{fi}(\mathbf{k}_f) \propto \big( A(\mathbf{k_{\parallel}}) \cdot M_A(\mathbf k_f) +  B(\mathbf{k_{\parallel}})  \cdot M_B(\mathbf k_f) \big) \times \\  \delta(h\nu - E_k - W - E_B),
\label{eq:IACA_methods}
   \end{aligned}
\end{equation} 
where $E_B$ is the binding energy, $W$ the work function, and $\mathbf k_f$ indicates the detected far field photoelectron with kinetic energy $E_k = (2m/\hbar)|\mathbf k_f |^2 $ and emitted along the direction $\mathbf k_f/ |\mathbf k_f |$.

With MS, the photoemission matrix element becomes
\begin{equation}
\begin{aligned}
M_{fi,MS}(\mathbf{k}_f) \propto \\ \big( A(\mathbf{k_{\parallel}}) \cdot M_{A,MS}(\mathbf k_f) +  B(\mathbf{k_{\parallel}})  \cdot M_{B,MS}( \mathbf k_f) \big) \times \\  \delta(h\nu - E_k - W - E_B),
\label{eq:MS_methods}
   \end{aligned}
\end{equation} 
where the patterns shown in Fig.~\ref{fig3:EDAC_MS}(a)-(d) need to be added coherently; thus, their amplitudes and phases, as well as different positions of emitters $A$ and $B$ are taken into account. 

\subsection{Details of calculation procedures}
Real-space calculations were performed using the EDAC photoelectron diffraction computer code \cite{Abajo2001}. For the atomic patterns shown in Fig.~\ref{fig1:CD-ARPES}(a)-(e) and Fig.~\ref{fig5:WSe2_atomic}(a)-(f) we used an inner potential of $V_0 = 0$ eV and muffin-tin potentials for the atoms in the cluster (touching muffin tins). For the radial wave function of C $2p$ we used values tabulated in Ref.~\cite{Clementi1974}, while for W $5d$ the radial wave function was calculated internally by EDAC. Spherical clusters centered at the emitter site were used. For W $5d$ we used the cluster radius of $R_{max} =  15$ Å, with 463 atoms spread through 3 layers of WSe$2$. For graphene we used $R_{max} =  20$ Å, cluster consisting 481 atoms. The inelastic mean free path of 5 Å was used in all cases. A surface barrier is simulated by the inner potential $V_0$, where we have used $V_0 = 17$ eV for grapehene and $V_0 = 13$ eV \cite{Riley2014} for WSe$_2$, however changing these values by a few eV does not change the calculated CD-ARPES maps considerably. Details on convergence of the EDAC calculations are provided in Supplementary Information.

One-step model calculations of graphene shown in  Fig.~\ref{fig4:omni} were performed using the $omni$ code with a setup used previously \cite{Gierz2011}. 

Our photoemission calculations do not take substrate-related effects into account.

Differential cross section (DCS) and Sherman function computation for the electron scattering on W atom have been performed using the ELSEPA code \cite{Salvat2021}. The results shown in Fig.~\ref{fig6:WSe2_normal_incidence} (h)-(j) have been calculated without taking into account correlation effects. Virtually the same results have been obtained with an atomic potential and with a muffin-tin potential with radius $R_{MT} = 1.35$ Å. Adding electron correlation on the level of the local density approximation (LDA), as implemented in ELSEPA, leads to small changes in the shapes of DCS and Sherman function, but does not qualitatively change the results.

\subsection{Sample preparation and photoemission measurements}
The measurements on graphene were performed at the NanoESCA beamline at Elettra using the modified Focus NanoESCA momentum microscope. The resolution was $\Delta E \approx 100$ meV and $\Delta k \approx 0.06$ Å$^{-1}$, the photon beam was incident at $65^\circ$ with respect to the surface normal. The spectrometer system allows probing the momentum space up to a radius of $\approx 2.5~$Å$^{-1}$ with respect to the Brillouin zone center. Samples were kept at $\approx 100K$. A dry transfer technique was used to prepare graphene/hBN heterostructures, the lateral sizes of the graphene flakes were $\approx 20 \mu$m, thus matching the photon beamspot of $\approx 20 \mu$m at NanoESCA. The twist angle between graphene and hBN was $\approx 20^\circ$. Details are provided in Supplementary Information.

WSe$_2$ and PtTe$_2$ measurements were performed on cleaved bulk single crystals at the PHELIX beamline at Solaris \cite{Szczepanik2021} at $\approx 40K$. The SPECS Phoibos 225 spectrometer resolution was set to $<50$ meV and all the maps were collected using the lens deflector system without rotating the sample during the measurement, thus probing the same sample area for each map. The endstation at PHELIX employs the magic angle of $54.7^\circ$ between the light incidence and the axis of the analyzer lens. Therefore, in the normal emission geometry, that is, when the surface normal coincides with the spectrometer lens axis, the light incidence is at $54.7^\circ$, while at normal light incidence the axis of the spectrometer lens is at $54.7^\circ$ off-normal. The electrostatic lens deflector system of the analyzer allows scanning $\pm 17^\circ$ angle in both directions without rotating the sample.

GdMn$_6$Sn$_6$ measurements were performed at the micro-ARPES branch of the MAESTRO beamline at the Advanced Light Source. Sample was cleaved by the ceramic post method, and the experiments were performed at $\approx 20K$, where the sample is magnetic state, however, with the beamspot of $>10 \mu m$ we have been averaging over magnetic domains.

\bibliography{Bibliography.bib}
 
\end{document}